\documentclass[aps,prb,preprint,showpacs,groupedaddress]{revtex4-1}  
\usepackage{graphicx}  
\usepackage{dcolumn}   
\usepackage{bm}        
\usepackage{amssymb}   

\begin{document}

\leftline{Version 11 of \today} 
\leftline{Primary author: Simon Hale}


\title{Effect of vertex corrections on the longitudinal transport through multilayered nanostructures: Exact dynamical mean-field theory approach applied to the inhomogeneous Falicov-Kimball model}
\author{S. T. F.~Hale$^{1}$}
\author{J. K. Freericks$^{1}$}
\affiliation{$^{1}$Department of Physics, Georgetown University, Washington, D.C 20057, USA}
\date{\today}

\begin{abstract}
Inhomogeneous dynamical mean-field theory is employed to calculate the vertex-corrected electronic charge transport for multilayered devices composed of semi-infinite metallic lead layers coupled through a strongly correlated material barrier region. The barrier region can be tuned from a metal to a Mott insulator through adjusting the interaction strength and the particle filling. We use the Falicov-Kimball model to describe the barrier region because an exact expression for the vertex corrections is known, allowing us to determine their effect on transport. The \textit{dc} conductivity is calculated and we find the effects of the vertex corrections are relatively small, manifesting themselves in a small reduction in the resistance-area product. This reduction saturates in absolute magnitude as the barrier layer becomes thick, as expected due to the vanishing nature of the vertex corrections in bulk. The vertex corrections have a larger relative effect on the resistance-area product for more metallic and thinner devices.
\end{abstract}

\maketitle 


\begin{center}
\textbf{I. INTRODUCTION}
\end{center}

	As the characteristic length scales of electronic devices reach further into the nanometer regime, quantum-mechanical effects give rise to novel properties and new devices. The field of strongly correlated materials makes possible the calculation of various properties of nanostructures. These calculations are motivated by the significant experimental activity in multilayered nano-heterostructures\cite{ohtomo, thiel, helmes}. New physical phenomena have been seen to emerge in these heterostructures that are absent in the bulk materials they are composed from, like the appearance of two-dimensional electron gases at the interfaces between band and Mott insulators{\cite{ohtomo}} and their low temperature superconductivity{\cite{thiel}}.

	In our work, we construct strongly correlated electron nanostructures, with carefully controlled quantum confinement effects. These theoretical nanostructure devices, are constructed from semi-infinite ballistic-metal leads that are coupled together through a barrier region of strongly correlated material.  We investigate the effect that the barrier region inhomogeneity has on the longitudinal charge transport through the multilayered nanostructures. Unlike previous work{\cite{freejk}}, that neglected the vertex corrections due to their vanishing nature in the bulk{\cite{khurana}}, we investigate the effects of including the vertex corrections in charge transport.  The vertex corrections vanish in the bulk due to the odd parity of the velocity operator in momentum space and the even parity of the local irreducible vertex{\cite{khurana}}. The arrangement of the multilayers breaks the homogeneity of the system meaning this parity argument no longer holds for longitudinal transport. Momentum is no longer a good quantum number in the longitudinal direction, so one cannot change the sign of the momentum and classify states according to their parity. Of course, for transport parallel to the planes, we have full translational invariance, so the vertex corrections do vanish in dynamical mean field theory. Recently, using different methods from our work, the importance of the vertex corrections in the optical conductivity calculations of the single-band Hubbard model using the dynamical cluster approximation has been investigated{\cite{linmillis}}, showing that as frequency and doping levels increase, the vertex corrections play an increasingly important role. Their technique does not require explicit calculation of the irreducible charge vertex, but instead they directly calculate the optical conductivity. Additional work by V. Jani\v{s} and V. Pokorn\'y{\cite{janic}} investigated general properties of the vertex corrections to the electrical conductivity of electrons scattered on random impurities. They found that the sign of the vertex corrections to the Drude conductivity is negative, which disagrees in sign, with the results presented in this paper, although the system they investigate is different from ours. Our work is concerned with the DC limit and investigates the effect the barrier thickness has on the vertex corrections. Due to the vanishing nature of the vertex corrections in the bulk, when the barrier is thick enough, it becomes bulk-like, so the vertex corrections stop having an effect, hence the maximal impact of the vertex corrections will be localized to the interface regions. 
	
\begin{figure}[htp]
	\centering
		\includegraphics[width=100mm]{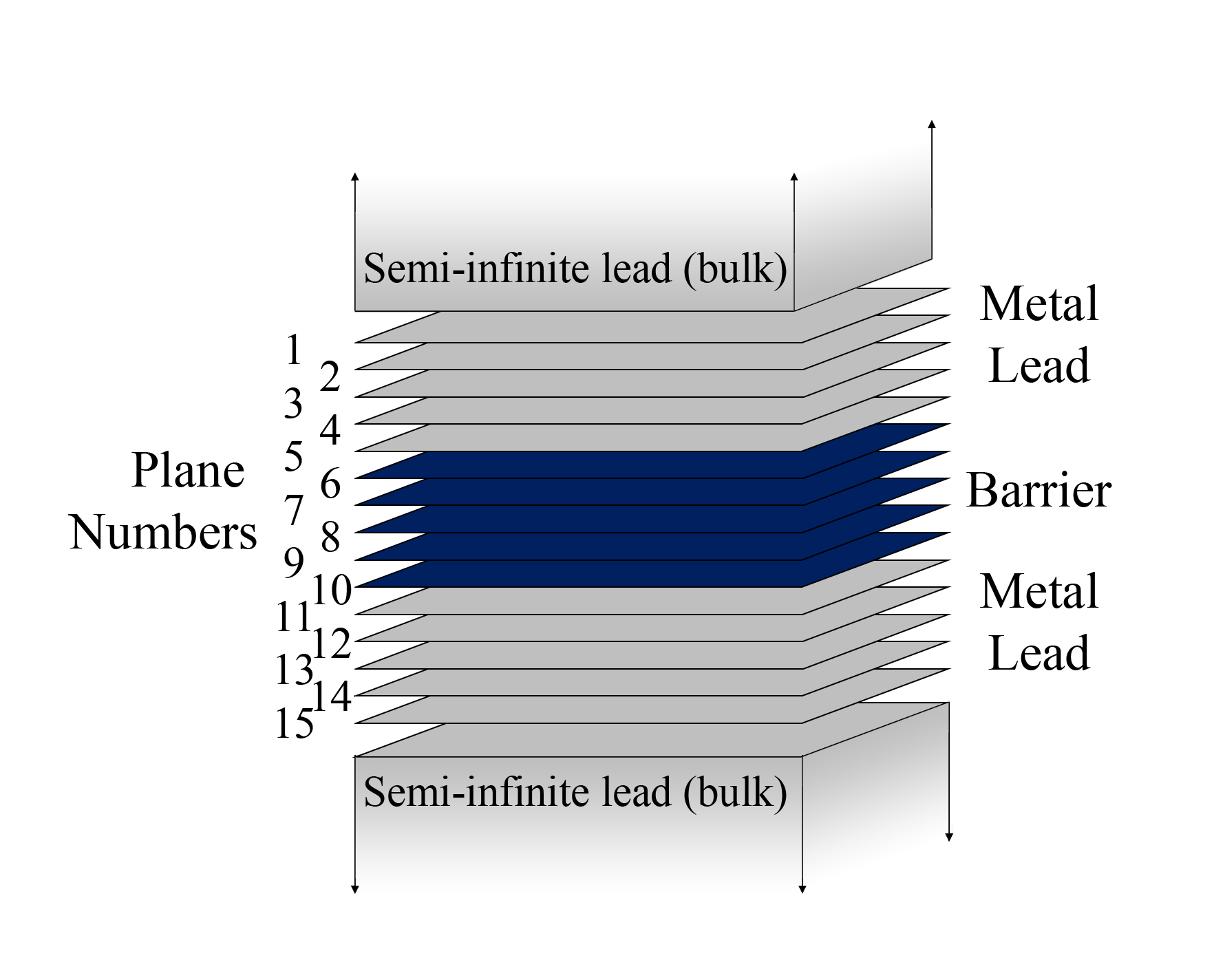}
	\caption{Schematic representation of the model, showing the stacking of homogeneous 2D layers forming a 3D multilayer nanostructure. The figure shows a metal-barrier-metal junction with only five metal layers on each side of the junction, although we can vary both the barrier and lead thicknesses. In the work presented here, the actual number of metal lead planes is 30 on each side and the number of barrier planes varies from one to fifty.}
\label{fig:layer}
\end{figure}

	The theoretical framework for this work is based on dynamical mean-field theory (DMFT) which allows for self-consistent calculations of the properties of strongly correlated materials. The DMFT formalism{\cite{pottnolt}} incorporates all forms of transport (except localization effects from weak disorder), therefore we don't have to make any \textit{a priori} assumptions about the type of transport in the device. We do make the assumption that the self energy is local within each plane, but can vary from plane to plane. For the interaction, we use the Falicov-Kimball model{\cite{Falkim}} and employ the Potthoff-Nolting{\cite{pottnolt}} technique of a mixed basis for the Green's functions in solving the inhomogeneous DMFT. This involves stacking two dimensional homogeneous $x-y$ planes in a longitudinal $z$-direction (see Fig.~\ref{fig:layer}), with hopping allowed only between neighboring planes in the $z$-direction. We allow for inhomogeneity in the $z$-direction, but preserve the translational invariance within each plane. This implies we can use momentum in the $x$ and $y$ directions, but we must solve the problem in real space in the $z$-direction. 
		
	 To begin, Fourier transform the $x$ and $y$ coordinates to wavevectors $k_x$ and $k_y$, respectively, but keep the $z$ coordinate in real space. In our formalism, we represent the $z$ coordinate, which indicates the plane numbers, with Greek letters ($\alpha$, $\beta$, $\gamma$,...). The Green's function then depends only on momentum through the two-dimensional band energy, so for each two-dimensional band energy, we solve a quasi-one-dimensional problem that is represented tridiagonally in real space and can be solved with the quantum zipper algorithm{\cite{freejk}}. We iterate our many-body equations to achieve a self-consistent solution.
	 
	In the remainder of this paper, we will work through the mathematical formalism in Sec. II. Section II also uses a Kubo formula for the electronic change transport including vertex corrections in DMFT to find the \textit{dc} conductivity matrix. We initially express the polarizability matrix on the imaginary-time axis, then we Fourier transform to the Matsubara frequencies, we take the analytic continuation to the real axis, and finally take the limit of the frequency as it goes to zero to find the \textit{dc}-conductivity matrix and the resistance. In Sec. III, we present our numerical results for various Falicov-Kimball interaction strengths and barrier layer thicknesses. Finally, in Sec. IV we give our conclusions and remark on possible future directions of research. 
\begin{center}
\textbf{II. FORMALISM}
\end{center}


To calculate the full electronic transport along the longitudinal direction (perpendicular to the planes), we use a Kubo-Greenwood-based linear-response formalism{\cite{Kubo, Greenwood}}. We operate in the steady state for current flow, with the charge density fixed as a function of time. A constant charge density implies that the charge current is conserved throughout the device because of the continuity equation. We also assume that temperature is a constant throughout the device and there is no electronic charge reconstruction{\cite{millis}} in our system, which is realistic when both the leads and the barrier are at half-filling and their chemical potentials match. The Kubo-Greenwood formula is based on the current-current correlation function of the charge current operator. The charge current operator for the $\alpha$th plane is the sum of all the current flowing longitudinally from the $\alpha$th plane to the $\alpha+1$st plane, 
\begin{equation}
\textbf{j}^{long}_{\alpha}  =  i a e t_{\alpha\alpha+1}\sum_{i\in plane}\left(c^{\dagger }_{\alpha+1 i}c_{\alpha i}-
c^{\dagger }_{\alpha i}c_{\alpha+1 i}\right)\hat{z},
\end{equation}
where $t_{\alpha\alpha+1}$ is the hopping between the $\alpha$th and $\alpha+1$st plane and $c^{\dagger }_{\alpha i}$ and $c_{\alpha i}$ are the creation and annihilation operators, respectively, for a conduction electron at lattice site $i$ on plane $\alpha$. Note that the index $i$ denotes the lattice site on each plane, a square lattice for this work, and the planes are all aligned in registry so the longitudinal hopping is straight across each plane from one equivalent site to the next equivalent site, i.e. the lattice sites are those of a simple cubic lattice. The totally current operator is then $\textbf{j}^{long}=\sum_{\alpha}\textbf{j}_\alpha^{long}$.

	We consider a Hamiltonian that involves a hopping term for the electrons and an interaction term for the sites within the
barrier region. For the interaction in our numerical calculations, we employ the Falicov-Kimball model{\cite{Falkim}} which involves an interaction between spinless conduction electrons and spinless localized electrons. When the conduction electron hops onto a site occupied by the localized electrons it feel a Coulomb repulsion. When this correlation strength is large enough, it has a Mott-like metal-insulator transition. In the second quantization formalism, the spinless Falicov-Kimball Hamiltonian{\cite{Falkim}} is, 
\begin{eqnarray}
H=-\sum_{\alpha}\sum_{i,j\in plane}t_{\alpha ij}c^{\dagger}_{\alpha i}c^{}_{\alpha j}
-\sum_{\alpha}\sum_{i\in plane}t_{\alpha \alpha+1}\left(c^{\dagger}_{\alpha i}c^{}_{\alpha+1 i}+c^{\dagger}_{\alpha+1 i}c^{}_{\alpha i}\right)\nonumber\\
-\mu \sum_{\alpha}\sum_{i\in plane} c^{\dagger}_{\alpha i}c^{}_{\alpha i}+
\sum_{\alpha}\sum_{i\in plane} U_{\alpha}c^{\dagger}_{\alpha i}c^{}_{\alpha i}\left(w_{\alpha i}-\frac{1}{2}\right),
\end{eqnarray}
where $t_{\alpha ij}$ is the intraplane hopping between nearest neighbor sites on plane $\alpha$, $U_{\alpha}$ represents the interaction strength on plane $\alpha$, and $w_{\alpha i}$ is a classical variable that equals one if there is a localized particle at site $i$ on plane $\alpha$ and zero if there is no localized particle at site $i$ on plane $\alpha$ (a chemical potential $\mu$ is employed to adjust the total conduction-electron concentration).

 It is important to note that the following derivation does not depend on the Hamiltonian used and can be thought of as a general result, until we explicitly state otherwise. We add the perturbation,
\begin{equation}
H'(t)=-\sum_{\alpha}\textbf{j}^{long}_{\alpha}\cdot\textbf{A}_{\alpha}(t),
\end{equation}
to the Hamiltonian, to introduce the effect of the electric field. $\textbf{A}(t)$ is the vector potential, with the electric field written in the gauge where the scalar potential vanishes, $\textbf{E}_{\alpha}=-\partial_{t}\textbf{A}_{\alpha}(t)$ and units chosen so $c=1$ (we also choose $\hbar=1$). The electric field, vector potential, and charge current all are vectors in the longitudinal ($z$) direction only.

To calculate the charge transport, we must start by evaluating correlation functions in imaginary time. To do so, we need to determine the Wick rotation for $H'(t)\rightarrow H'(\tau)$. Since the vector potential couples to the charge current, we take the vector potential on the imaginary time axis to be periodic in $\beta=1/T$, so it can be expressed as a Fourier series in the bonsonic Matsubara frequencies $A_{\alpha}(\tau)=T \sum_{l} \exp [i \nu_{l} \tau] A_{\alpha l}$ with $i \nu_{l}=2 i \pi T l$.
The current-current correlation function for the multilayer nanostructure is defined to be
\begin{equation}
\Pi_{\alpha\beta}(iv_{l}) = \int^{\beta}_{0}d\tau e^{iv_{l}\tau}\left\langle T_{\tau} \textbf{j}^{\dagger long}_{\alpha}(\tau) \textbf{j}^{long}_{\beta}(0)\right\rangle,
\label{eq:PIcc}
\end{equation}
with $\alpha$ and $\beta$ being planar indices (not to be confused with the $\beta$ in the integration limit which is the inverse temperature) and $T_{\tau}$ denotes time ordering of operators. The notation $\left\langle X \right\rangle$ denotes the trace, ${\rm Tr }\  \exp(-\beta H) X$ divided by the partition function $\cal{Z}$$ ={\rm Tr }\   \exp(-\beta H )$, and the operators are expressed in the Heisenberg representation $X(\tau)=\exp(\tau H) \  X \  \exp(-\tau H)$, all with respect to the equilibrium Hamiltonian $H$.

We calculate the current-current correlation function by taking the functional derivative of the appropriate expectation value with respect to the vector potential. We will use Green's functions to express our results. Because we need to know the dependence of the Green's function on $A$, we need to define a two-time Green's function in the presence of $H+H'(\tau)$.
The Green's function, in real space, is defined by
\begin{equation}
G_{\alpha\beta ij}(\tau,\tau';A)=-\left\langle T_{\tau} c_{\alpha i}(\tau;A) c^{\dagger}_{\beta j} (\tau';A) \right\rangle,
\label{eq:greens}
\end{equation}
for imaginary time $\tau$ and $\tau'$. The extra notation ``$;A$'' is used to remind us that the operators $c_{i \alpha}$ and $c^{\dagger}_{j \beta}$ evolve according to the full Hamiltonian $H+H'(\tau)$ via $\hat{O}(\tau;A)=U^{\dagger}(\tau, 0)\  \hat{O} \ U(\tau, 0)$ with $U(\tau, 0)$ being the evolution operator with respect to $H+H'(\tau)$; i.e. $\partial_{\tau} U(\tau, 0)=[H+H'(\tau)] U(\tau,0)$. To properly express the Green's functions for the Matsubara  frequencies we use a double Fourier transformation
\begin{equation}
G_{\alpha\beta}(i\omega_{n},i\omega_{m};A)=T \int^{\beta}_{0}d\tau \int^{\beta}_{0}d\tau' e^{i\omega_{n}\tau} G_{\alpha\beta}(\tau,\tau';A)e^{-i\omega_{m}\tau'}.
\label{eq:GInt}
\end{equation}
We do this because the perturbation $\textbf{j}\cdot \textbf{A}(\tau)$ causes the Green's function to lose its time-translational invariance. Note that after taking all relevant derivatives and algebraic manipulations, we can express the Green's functions in zero field, where $G_{\alpha\beta}(i\omega_{n},i\omega_{m};A)\propto \delta_{nm}$, and we write $G_{\alpha\beta}(i\omega_{n})$ because the limiting form of the Green's function in Eq.~(\ref{eq:greens}) depends only on $\tau-\tau'$ in equilibrium as $A(\tau)\rightarrow 0$.

	To build our model, we need to solve for the local Green's function on each plane, which we do by employing the quantum ``zipper'' algorithm{\cite{freejk}}, based on the Potthoff-Nolting formalism{\cite{pottnolt}}. When solving for the local Green's functions, we use the symbol $Z$ to represent a general variable in the complex plane. Typically $Z$ takes the form of the fermionic Matsubara frequencies, $Z=i \omega_{n}$ or the analytic continuation to the real axis with $i \omega_{n}\rightarrow\omega\pm i 0^{+}$. We start with the unperturbed equation of motion (EOM), for equilibrium, where $A_{\alpha}=0$,
\begin{eqnarray}
\sum_{\gamma}G_{\gamma\beta}(Z;\bm{k}^{||})\left[ \left( Z+\mu-\epsilon^{\alpha}_{\bm{k}^{||}}\right) \delta_{\gamma\alpha}+\left(t_{\alpha-1\alpha}\delta_{\gamma\alpha-1}+t_{\alpha+1\alpha}\delta_{\gamma\alpha+1}\right)
-\Sigma_{\alpha}(Z)\delta_{\gamma\alpha}\right]=\delta_{\alpha\beta},
\end{eqnarray}
where the two-dimensional band structure is $\epsilon^{\alpha}_{\bm{k^{||}}}=-2t_{\alpha}\left[\cos k_{x}+\cos k_{y}\right]$, $\bm{k^{||}}=(k_{x},k_{y},0)$ is the transverse momentum, $\delta_{\alpha\beta}$ is the Kronecker delta function, and $\Sigma_{\alpha}(Z)$ is the local self-energy on plane $\alpha$. Note that from this point on we will use the simplification that the hopping matrix elements are equal to $t$ for all nearest neighbors (interplane and intraplane), $t_{\alpha+1\alpha}=t_{\alpha-1\alpha}=t_{\alpha i j}=t$ and vanish otherwise. Since the EOM has a tridiagonal form with respect to the spatial component, it can be solved with the so-called renormalized perturbation expansion\cite{econ}. We solve the equation directly, for the $\beta=\alpha$ case via
\begin{equation}
G_{\alpha\alpha}(Z;\bm{k}^{||})=\frac{1}{Z+\mu-\Sigma_{\alpha}(Z)-\epsilon_{\bm{k}^{||}\alpha}+\frac{G_{\alpha-1\alpha}(Z;\bm{k}^{||})}{G_{\alpha\alpha}(Z;\bm{k}^{||})}t
+\frac{G_{\alpha\alpha+1}(Z;\bm{k}^{||})}{G_{\alpha\alpha}(Z;\bm{k}^{||})}t}.
\end{equation}
We create left and right recursion relations, 
\begin{equation}
L_{\alpha-n}(Z;\bm{k}^{||})=Z+\mu-\Sigma_{\alpha-n}(Z)-\epsilon_{\bm{k}^{||}}+\frac{t^{2}}{L_{\alpha-n-1}(Z;\bm{k}^{||})},
\end{equation}
and 
\begin{equation}
R_{\alpha+n}(Z;\bm{k}^{||})=Z+\mu-\Sigma_{\alpha+n}(Z)-\epsilon_{\bm{k}^{||}}+\frac{t^{2}}{R_{\alpha+n+1}(Z;\bm{k}^{||})}
\end{equation}
respectively. We start these relationships with the bulk values ($n\rightarrow \pm \infty$), which give us 
\begin{equation}
L_{-\infty}(Z;\bm{k}^{||})=\frac{Z+\mu-\Sigma_{-\infty}(Z)-\epsilon_{\bm{k}^{||}}}{2}\pm \sqrt{[Z+\mu-\Sigma_{-\infty}(Z)-\epsilon_{\bm{k}^{||}}]^{2}-4t^{2}}
\end{equation}
and
\begin{equation}
R_{\infty}(Z;\bm{k}^{||})=\frac{Z+\mu-\Sigma_{\infty}(Z)-\epsilon_{\bm{k}^{||}}}{2}\pm \sqrt{[Z+\mu-\Sigma_{\infty}(Z)-\epsilon_{\bm{k}^{||}}]^{2}-4t^{2}}.
\end{equation}
The signs in the previous two equations are chosen to yield an imaginary part less than zero for $Z$ lying in the upper half plane, and vice versa for $Z$ lying in the lower half plane. The self-energies ($\Sigma_{\pm\infty}$) vanish for the ballistic metal leads used here. Employing these recursive relationships, we can solve the diagonal, $\alpha=\beta$, and off-diagonal terms, ($\alpha\neq\beta$), using the relations
\begin{equation}
G_{\alpha\alpha-n}(Z;\bm{k}^{||})=-\frac{G_{\alpha\alpha-n+1}(Z;\bm{k}^{||})t}{L_{\alpha-n}(Z;\bm{k}^{||})}
\label{eq:offdiag1}
\end{equation}
and
\begin{equation}
G_{\alpha\alpha+n}(Z;\bm{k}^{||})=-\frac{G_{\alpha\alpha+n-1}(Z;\bm{k}^{||})t}{R_{\alpha+n}(Z;\bm{k}^{||})},
\label{eq:offdiag2}
\end{equation}
which are defined for $n>0$. In addition, one needs to note that the following identity holds: $G_{\alpha \beta}(Z;\bm{k}^{||})=G_{\beta \alpha}(Z;\bm{k}^{||})$, by interchanging planar subscripts in the EOM, solving for the new recursion relationships, and showing they are the same as the above.

We need to express the current-current correlation function in a form where we can use the local Green's functions for each plane. We begin by using Green's functions to evaluate the expectation value of the current-current correlation function in Eq.~(\ref{eq:PIcc}),
\begin{eqnarray}
\Pi_{\alpha\beta}(iv_{l})=\sum_{mn}\sum_{\bm{k}^{||}}i a e t\left[\frac{\delta G_{\beta\beta+1}\left(i\omega_{m},i\omega_{n};\bm{k}^{||};A\right) }{\delta A_{\alpha,-l}} - \frac{\delta G_{\beta+1\beta}\left(i\omega_{m},i\omega_{n};\bm{k}^{||};A\right)}{\delta A_{\alpha,-l}} \right],
\end{eqnarray} 
where $\delta$ denotes a functional derivative. After employing the identity 
\begin{eqnarray}
G_{\alpha\beta}(i\omega_{n},i\omega_{m};A)= \sum_{m'n'}\sum_{\gamma\delta} G_{\alpha\gamma}(i\omega_{n},i\omega_{m'};A)G^{-1}_{\gamma\delta}(i\omega_{m'},i\omega_{n'};A)G_{\delta\beta}(i\omega_{n'},i\omega_{m};A),
\end{eqnarray}
we have
\begin{eqnarray}
\Pi_{\alpha\beta}(iv_{l})=\sum_{mnm'n'}\sum_{\bm{k}^{||}}\sum_{\gamma\delta}i a e t\frac{\delta G^{-1}_{\gamma\delta}\left(i\omega_{m'},i\omega_{n'};\bm{k}^{||};A\right) }{\delta A_{\alpha,-l}}\nonumber\\
\left[ G_{\beta\gamma}\left(i\omega_{m};\bm{k}^{||}\right) 
  G_{\delta\beta+1}\left(i\omega_{n};\bm{k}^{||}\right)\right. \nonumber\\
\left.- G_{\beta+1\gamma}\left(i\omega_{m};\bm{k}^{||}\right) 
  G_{\delta\beta}\left(i\omega_{n};\bm{k}^{||}\right)\right]\delta_{m m'}\delta_{n n'}.
\end{eqnarray}
Note that we replaced Green's functions that have no derivative with respect to the vector potential by their zero field values as is done in the Kubo-Greenwood formula.
	
	Calculating the functional derivative of the inverse of the Green's function is easier then calculating the functional derivative of the Green's function itself. We find the inverse of the Green's function by examining the EOM for the Green's function in a field, a generalization of Eq.~(\ref{eq:GInt}) to include two-time dependence and the vector potential, yielding 
\begin{eqnarray}
\sum_{\gamma}\sum_{m'}G_{\gamma\delta}(i\omega_{m'},i\omega_{n};\bm{k}^{||};A)\left[ \left( i\omega_{m}+\mu-\epsilon^{\alpha}_{\bm{k}^{||}}\right) \delta_{\alpha\gamma}\delta_{mm'}
+\left(t\delta_{\gamma\alpha-1}+t\delta_{\gamma\alpha+1}\right)\delta_{mm'}\right. \nonumber\\
-\Sigma_{\alpha}(i\omega_{m},i\omega_{m'})\delta_{\alpha\gamma}+iaetT A_{\alpha,l}\delta_{\gamma\alpha+1}\delta_{m'm+l}
\left.-iaetT A_{\alpha-1,-l}\delta_{\gamma\alpha-1}\delta_{m'm+l}\right]=\delta_{\alpha\delta},
\label{eq:eom}
\end{eqnarray}
with  $G^{-1}(i\omega_{m},i\omega_{m'};\bm{k}^{||};A)$ being the terms inside the brackets.
Taking the functional derivative of $G^{-1}(i\omega_{m},i\omega_{m'};\bm{k}^{||};A)$ with respect to $A_{\alpha,-l}$, only three terms have nonzero derivative; resulting in
\begin{eqnarray}
\frac{\delta G^{-1}_{\gamma\delta}\left(i\omega_{m'},i\omega_{n'};\bm{k}^{||};A\right) }{\delta A_{\alpha,-l}}&=&-\sum_{m''n''}\frac{\delta\Sigma_{\gamma}\left(i\omega_{m'},i\omega_{n'};A\right) }{\delta G_{\gamma\gamma}\left(i\omega_{m''},i\omega_{n''};A\right)}
\frac{\delta G_{\gamma\gamma}\left(i\omega_{m''},i\omega_{n''};A\right) }{\delta A^{\alpha}_{-l}}
\nonumber\\
&&+iaeT\left(t\delta_{\delta\gamma+1}\delta_{\alpha\gamma}
-t\delta_{\delta\gamma-1}\delta_{\alpha+1\gamma}\right)\delta_{m'+l,n'}.
\label{eq:dgda}
\end{eqnarray}
Evaluating the Green's functions in zero field [$G\left(i\omega_{n'},i\omega_{n};\bm{k}^{||}\right)\propto \delta_{nm}$]
and noting that any Green's function that does not have a derivative acting on it can be replaced by it's (diagonal) zero-field value, the current-current correlation function becomes
\begin{eqnarray}
\Pi_{\alpha\beta}(iv_{l})&=&
a^{2}e^{2}t^{2}T\sum_{m}\sum_{\bm{k}^{||}}\left[ 
G_{\beta\alpha}\left(i\omega_{m};\bm{k}^{||}\right) 
  G_{\alpha+1\beta+1}\left(i\omega_{m+l};\bm{k}^{||}\right)\right. 
\nonumber\\
& &+G_{\beta+1\alpha+1}\left(i\omega_{m};\bm{k}^{||}\right) 
  G_{\alpha\beta}\left(i\omega_{m+l};\bm{k}^{||}\right)
-G_{\beta\alpha+1}\left(i\omega_{m};\bm{k}^{||}\right) 
  G_{\alpha\beta+1}\left(i\omega_{m+l};\bm{k}^{||}\right)
  \nonumber\\
& &\left.-G_{\beta+1\alpha}\left(i\omega_{m};\bm{k}^{||}\right) 
  G_{\alpha+1\beta}\left(i\omega_{m+l};\bm{k}^{||}\right)\right]
\nonumber\\
&+&iaet\sum_{mnm''n''}\sum_{\bm{k}^{||}}\sum_{\gamma\delta}
\left[ G_{\beta\gamma}\left(i\omega_{m};\bm{k}^{||}\right) 
  G_{\gamma\beta+1}\left(i\omega_{n};\bm{k}^{||}\right)\right. 
\nonumber\\
& &\left.- G_{\beta+1\gamma}\left(i\omega_{m};\bm{k}^{||}\right) 
  G_{\gamma\beta}\left(i\omega_{n};\bm{k}^{||}\right)\right]
\frac{\delta\Sigma_{\gamma}\left(i\omega_{m},i\omega_{n};A\right) }{\delta G_{\gamma\gamma}\left(i\omega_{m''},i\omega_{n''};A\right)}
\frac{\delta G_{\gamma\gamma}\left(i\omega_{m''},i\omega_{n''};A\right) }{\delta A_{\alpha,-l}}.
\label{eq:presub}
\end{eqnarray}
The second term on the right hand side is the vertex correction which disappears in the bulk due to parity arguments, but here is nonzero. The functional derivative of $\Sigma$ with respect to $G$ is proportional to the irreducible charge vertex. 

	It is worth nothing again that up until this point we have not made use of the Falicov-Kimball model. Hence, the formalism is independent of the model used (periodic Anderson, Falicov-Kimball, Hubbard, etc.). The Falicov-Kimball model is now explicitly used because the result for the irreducible charge vertex $\Gamma$ is known to be{\cite{shvmfree}}, 
\begin{equation}
\Gamma_{\gamma}\left(i\omega_{m},i\omega_{n}\right)=\frac{1}{T}\frac{\delta\Sigma_{\gamma}\left(i\omega_{m},i\omega_{n};A\right) }{\delta G_{\gamma}\left(i\omega_{m''},i\omega_{n''};A\right)}
=\frac{1}{T}\frac{\Sigma_{\gamma}(i\omega_{m})-\Sigma_{\gamma}(i\omega_{n})}{ G_{\gamma}(i\omega_{m})-G_{\gamma}(i\omega_{n})}\delta_{mm''}\delta_{nn''},
\end{equation}
in the limit of vanishing field. The delta functions on the frequencies are specific to the Falicov-Kimball model, other models will have more complicated formulas than what we develop below.

	Substituting this expression and Eq.~(\ref{eq:dgda}) into Eq.~(\ref{eq:presub}) we are left with the full equation for the current-current correlation function on the imaginary frequency axis:
\begin{eqnarray}
\Pi_{\alpha\beta}(iv_{l})&=&
e^{2}a^{2}t^{2}T\sum_{m}\left\{\sum_{\bm{k}^{||}}\left[ 
G_{\beta\alpha}\left(i\omega_{m};\bm{k}^{||}\right) 
  G_{\alpha+1\beta+1}\left(i\omega_{m+l};\bm{k}^{||}\right)\right.\right.
\nonumber\\
& &+G_{\beta+1\alpha+1}\left(i\omega_{m};\bm{k}^{||}\right) 
  G_{\alpha\beta}\left(i\omega_{m+l};\bm{k}^{||}\right)
-G_{\beta\alpha+1}\left(i\omega_{m};\bm{k}^{||}\right) 
  G_{\alpha\beta+1}\left(i\omega_{m+l};\bm{k}^{||}\right)
  \nonumber\\
& &\left.-G_{\beta+1\alpha}\left(i\omega_{m};\bm{k}^{||}\right) 
  G_{\alpha+1\beta}\left(i\omega_{m+l};\bm{k}^{||}\right)\right]
\nonumber\\
&+&\sum_{\gamma\delta}\left(\sum_{\bm{k}^{||}}
\left[ G_{\beta\gamma}\left(i\omega_{m};\bm{k}^{||}\right) 
  G_{\gamma\beta+1}\left(i\omega_{m+l};\bm{k}^{||}\right) 
- G_{\beta+1\gamma}\left(i\omega_{m};\bm{k}^{||}\right) 
  G_{\gamma\beta}\left(i\omega_{m+l};\bm{k}^{||}\right)\right]\right.
\nonumber\\
&\times &\frac{\Sigma_{\gamma}(i\omega_{m})-\Sigma_{\gamma}(i\omega_{m+l})}{G_{\gamma}(i\omega_{m})-G_{\gamma}(i\omega_{m+l})}
\nonumber\\
&\times &\left[\delta_{\gamma\delta}-\sum_{\bar{\bm{k}}^{||}}G_{\gamma\delta}(i\omega_{m};\bar{\bm{k}}^{||}) 
\frac{\Sigma_{\delta}(i\omega_{m})-\Sigma_{\delta}(i\omega_{m+l})}{G_{\delta}(i\omega_{m})-G_{\delta}(i\omega_{m+l})}
G_{\delta\gamma}(i\omega_{n};\bar{\bm{k}}^{||})\right]^{-1}_{\gamma\delta}
\nonumber\\
&\times &\left.\left.\sum_{\bar{\bar{\bm{k}}}^{||}}\left[G_{\delta\alpha}(i\omega_{m};\bar{\bar{\bm{k}}}^{||}) G_{\alpha+1\delta}(i\omega_{m+l};\bar{\bar{\bm{k}}}^{||})- G_{\delta\alpha+1}(i\omega_{m};\bar{\bar{\bm{\bm{k}}}}^{||}) G_{\alpha\delta}(i\omega_{m+l};\bar{\bar{\bm{k}}}^{||})\right]\right)\right\},
\end{eqnarray}
where $\bm{k}^{||}, \bar{\bm{k}}^{||}, \bar{\bar{\bm{k}}}^{||}$ are all transverse momenta in two-dimensions and the inverse denotes a matrix inversion with respect to the planar indices $\gamma,\delta$ of the matrix defined in the brackets.

	To calculate the electronic charge transport we need the real-axis response. This is found by performing an analytic continuation from the imaginary to the real frequency axis{\cite{sharvin}}. We rewrite the Matsubara summations using contour integrations in the standard way that encircles all the simple poles corresponding to the Matsubara frequencies. The contour integrals have contributions at the poles of the Fermi-Dirac distribution [$f(\omega)=1/(1+\exp (\beta\omega))$] which lie at the fermionic Matsubara frequencies, $i \omega_{n}=i\pi T(2n+1)$ (the residue of the pole is $-T$). The contours are then deformed to lines parallel to the real axis and the Green's functions are evaluated with either retarded or advanced functions depending on the argument and the regions of analyticity (since the functions are all analytic, there are no additional poles). We then set the Fermi-Dirac function at $\omega-iv_{l}$ equal to $f(\omega)$ before analytical continuing the $iv_{l}$ frequency to the real axis. Finally, we continue $iv_{l}$ to $v+i \delta$ and shift the $\omega+v\rightarrow \omega $ in relevant integrals to obtain the real axis current-current correlation function. 
 
To find the \textit{dc} conductivity, we need to take the limit as the frequency $v$ goes to zero of the polarization, 
\begin{equation}
\sigma_{\alpha\beta}(0)=\lim_{v \to 0}\mbox{Re} \frac{i\prod_{\alpha\beta}(v)}{v}.
\end{equation}
The final result is obtained after a lengthy algebraic manipulation focused on rearranging the $\alpha$ and $\beta$ labels and explicitly writing out the real and imaginary parts, $\rm {Re}(Z)=\frac{Z+Z^{*}}{2}$ and $\rm {Im}(Z)=\frac{Z-Z^{*}}{2i}$, and expanding the paired terms.
We also use the fact that $G_{\alpha\beta}=G_{\beta\alpha}$ to arrive at our final expression for the \textit{dc}-conductivity,
\begin{eqnarray}
\sigma_{\alpha\beta}(0)&=&\frac{e^{2}a^{2}t^{2}}{\pi} \int d\omega 
\left(-\frac{\partial f(\omega)}{\partial \omega}\right) \Biggl\{\sum_{\bm{k}^{||}}
\left[{\rm Im} G_{\beta\alpha}(\omega;\bm{k}^{||}){\rm Im} G_{\alpha+1\beta+1}(\omega;\bm{k}^{||})\right.
\nonumber\\
& &+{\rm Im} G_{\beta+1\alpha+1}(\omega;\bm{k}^{||}){\rm Im} G_{\alpha\beta}(\omega;\bm{k}^{||})
-{\rm Im} G_{\beta\alpha+1}(\omega;\bm{k}^{||}){\rm Im} G_{\alpha\beta+1}(\omega;\bm{k}^{||})
\nonumber\\
& &\left.-{\rm Im} G_{\beta+1\alpha}(\omega;\bm{k}^{||}){\rm Im} G_{\alpha+1\beta}(\omega;\bm{k}^{||})\right]
\nonumber\\
&+&\sum_{\bm{k}^{||}} {\rm Im} \left[G^{*}_{\beta\gamma}(\omega;\bm{k}^{||}) 
  G_{\gamma\beta+1}(\omega;\bm{k}^{||})\right]
\nonumber\\
& &\times\frac{{\rm Im}\Sigma_{\gamma}(\omega)}{{\rm Im} G_{\gamma}(\omega)}
\left[\delta_{\gamma\delta}-\sum_{\bar{\bm{k}}^{||}}G^{*}_{\gamma\delta}(\omega;\bar{k}^{||}) 
\frac{{\rm Im}\Sigma_{\delta}(\omega)}{{\rm Im} G_{\delta}(\omega)}
G_{\delta\gamma}(\omega;\bar{\bm{k}}^{||})\right]^{-1}_{\gamma\delta}
\nonumber\\
& &\times\left.\sum_{\bar{\bar{\bm{k}}}^{||}}{\rm Im} \left[ G^{*}_{\delta\alpha}(\omega;\bar{\bar{\bm{k}}}^{||})G_{\alpha+1\delta}(\omega;\bar{\bar{\bm{k}}}^{||}) \right] \right\}.
\label{eq:dccond}
\end{eqnarray}
In numerical calculations, since the $\bm{k}$ dependence is always through $\epsilon^{2d}_{\bm{k}}$, we replace all momentum summations by integrals over the 2-d DOS. Note that one can also arrive at Eq.~(\ref{eq:dccond}) by using the general analytic continuation formulas in Ref. \onlinecite{FreeZS}.

	Given the conductivity matrix, we can extract the resistance of the multilayer nanostructure. This is done by relating the electric field and the expectation value of the planar current density in linear response and utilizing Ohm's law, giving the resistance-area per unit cell product of the nanostructure
\begin{equation}
R_{n}a^{2}=\sum_{\alpha\beta}[\sigma(0)]^{-1}_{\alpha\beta}.
\end{equation}
Note that it is the resistance $R_{n}$ that is extracted from these calculations, not the resistivity. Since the system is inhomogeneous, we do not have the appropriate geometrical factors to convert resistance into the resistivity as one can do in a homogeneous system. 

	The full algorithm for our work using dynamical mean-field theory can now be summarized. We start with the self-consistent iterative algorithm outlined in Ref. \onlinecite{freejk} to calculate the self-energies and diagonal Green's function on all planes ($G_{\alpha\alpha})$. We use a similar parallel approach exploiting the fact that the frequencies are completely independent of each other, to calculate the Green's functions at different frequencies on different processors.  After calculating the diagonal Green's functions, we use Eqs.~(\ref{eq:offdiag1}) and ~(\ref{eq:offdiag2}) to calculate all of the off-diagonal Green's functions. Once all of the Green's functions have been calculated we can use them in Eq.~(\ref{eq:dccond}), noting that at half filling, the temperature dependence only enters into the Fermi factor derivative, allowing for us to calculate the temperature dependence of $\sigma_{\alpha\beta}(0)$ without having to recalculate the Green's functions, though all numerical results presented here will be at one temperature.

\begin{center}
\textbf{III. RESULTS}
\end{center}

	We perform our calculations at half-filling ($\mu=0,\langle c^{\dagger}_{i}c_{i} \rangle=1/2$, and $w_{1}=\langle w_{i} \rangle =1/2$). This has a number of advantages. First, because the chemical potential is the same for the metallic leads and the barrier, the filling remains homogeneous throughout the system and there is no electronic charge reconstruction. Second, the chemical potential is fixed as a function of temperature ($\mu=0$), so we don't have to recalculate the chemical potential as the temperature is changed.
	We carry out our calculations on a simple cubic lattice allowing for nearest neighbor hopping only (interplane and intraplane hopping are equal). This allows us to reduce the number of parameters that we vary for our calculations and focus on the physical properties. In our calculations, we also include 30 self-consistent planes in the metallic leads both to right and left of the barrier, which we vary in thickness between 1 and 50 planes.
\begin{figure}[htp]
	\centering
		\includegraphics[width=80mm]{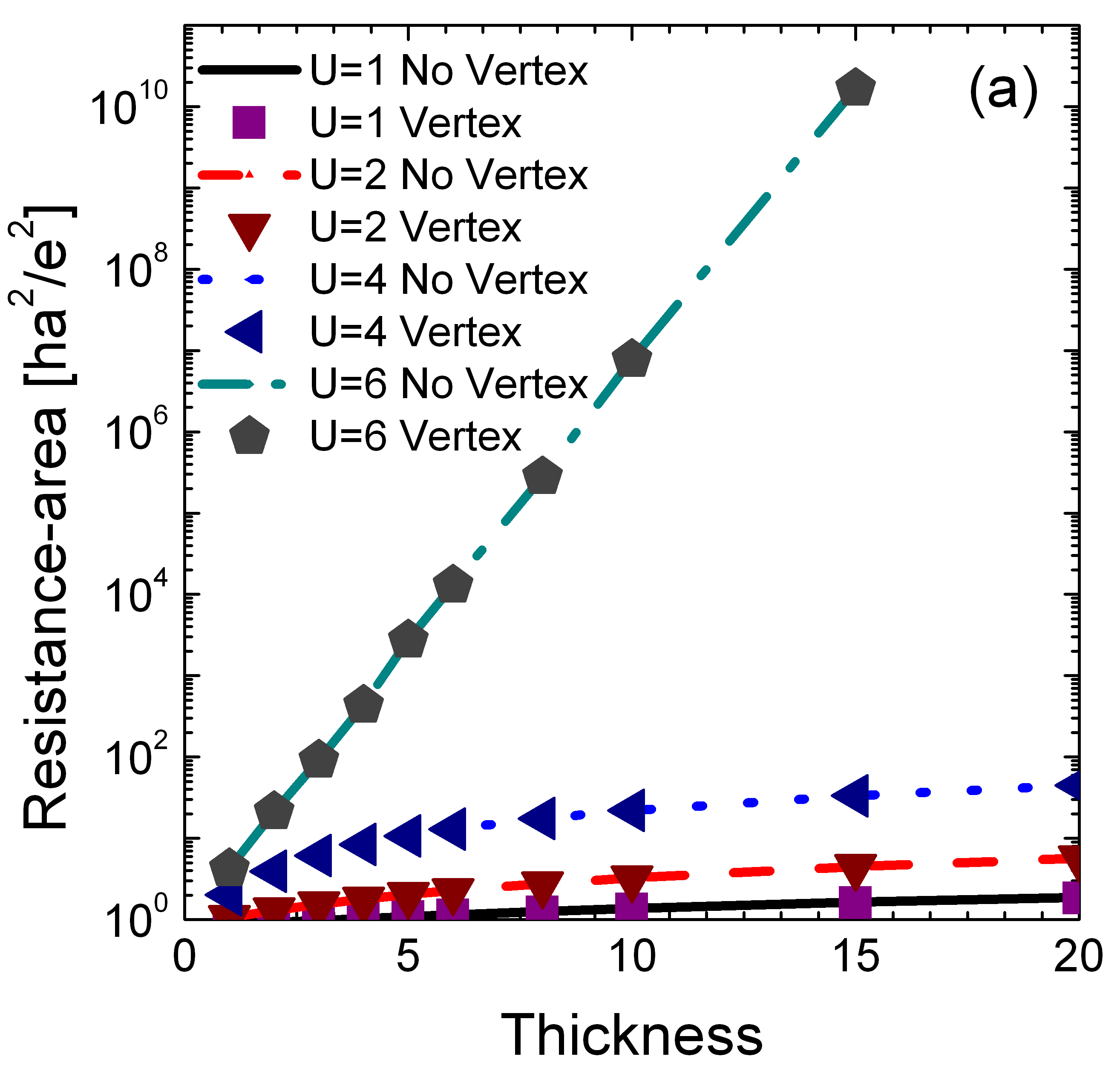}
		\includegraphics[width=80mm]{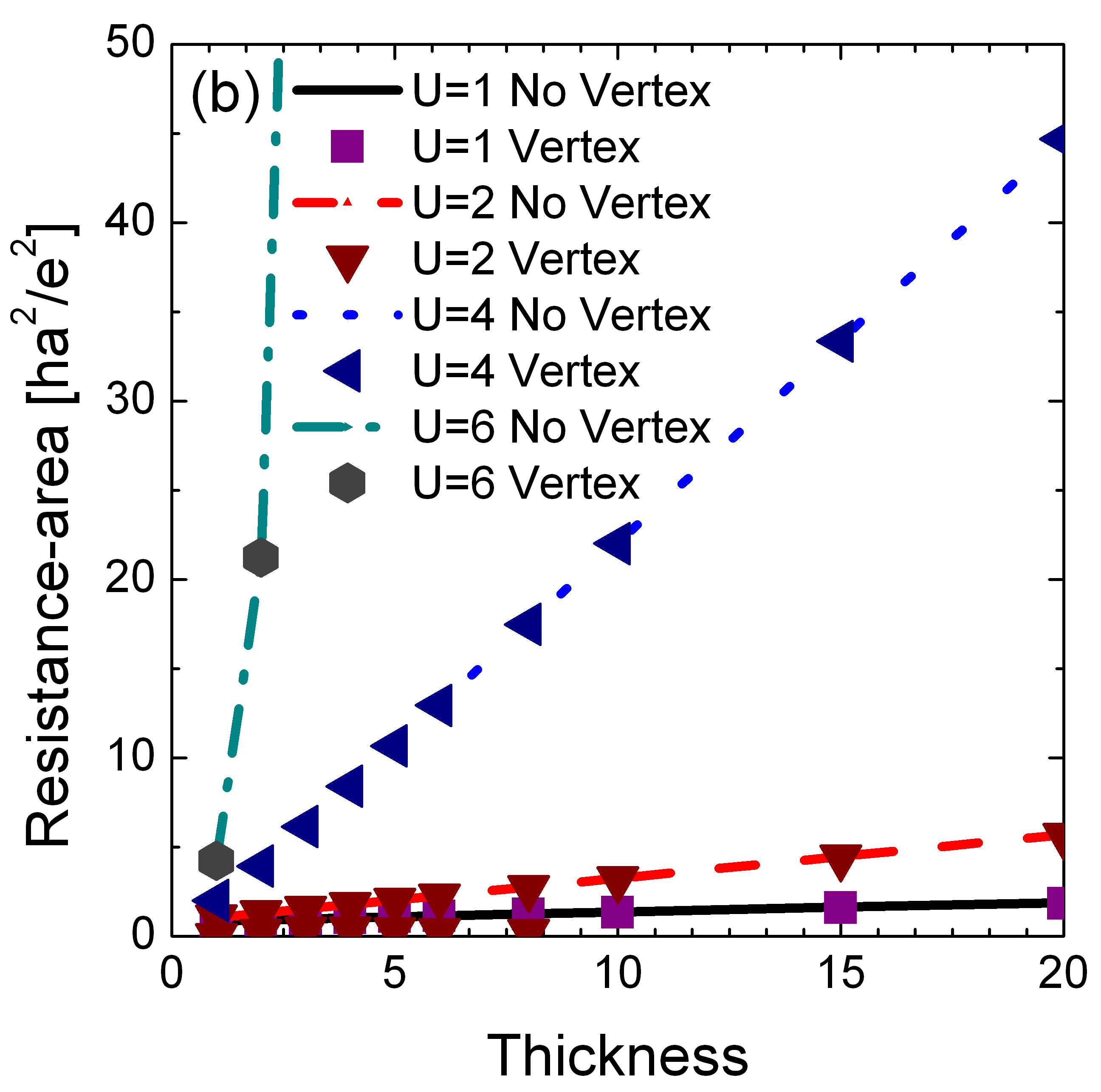}
	\caption{(Color online) Resistance-area product for nanostructures with $U=1,2,4,$ and $6$, and various thicknesses. Panel (a) is a semilogarithmic plot, while panel (b) is a linear plot. The temperature is $T=0.01$ in both panels. Note how the correlated insulator ($U=6$) has an exponential growth with thickness as expected{\cite{freejk}} for a tunneling process at low enough temperature, while the metallic cases ($U=1,2$, and 4) have nearly perfect linear scaling. The values with (symbols) and without (lines) the vertex correction are indistinguishable on the scale of the graphs.}
		\label{fig:layers}
\end{figure}  	
	
	We vary the Falicov-Kimball interaction strength ($U$) through its metal-insulator transition, which for our simple cubic lattice occurs at $U=2\sqrt{6}\approx4.92$. We plot the resistance-area product versus thickness for four different U values (Fig. \ref{fig:layers}): $U=1$ and $U=2$, in the diffusive metal regime; $U=4$, a strongly scattering, anomalous metal; $U=6$, a Mott insulator with a small correlation-induced gap. These plots are both for vertex-corrected and non vertex-corrected expressions for the conductivity matrices. The plots in Fig. \ref{fig:layers} show the expected behavior; for $U=1$, $U=2$, and $U=4$, all show an Ohm's law linear scaling of resistance with thickness, characteristic of diffusive metals, with a nonzero intercept denoting the non-vanishing contact resistance associated with the metallic leads. Additionally, the resistance-area product for the Mott insulator, $U=6$ grows exponentially with thickness due to its tunneling behavior. Note that the vertex corrections hardly have any effect on the behavior in either regime. 

	There are a number of numerical details that need to be discussed. We need to ensure that at low energies the properties of the nanostructures are accurately determined, which becomes increasingly difficult as we examine thicker Mott insulators and therefore requires more computational power. Due to these limitations, we limit our thickness for various interaction strengths, $U=1$ up to 50 layers, $U=2$ up to 30 layers, $U=4$ up to 20 layers, and $U=6$ up to 15 layers. To accurately calculate the sharp frequency dependence of the self-energy, we use a step size as fine as 0.001 for the frequency grid and up to half a million points for the integration over the two-dimensional DOS. The combination of the exponential growth of the resistance-area product when $U=6$ and the inherent numerical accuracy we achieve makes it difficult to extract the difference between vertex corrected and non vertex corrected results for thicker barriers.

	\begin{figure}[htp]
	\centering
				\includegraphics[width=80mm]{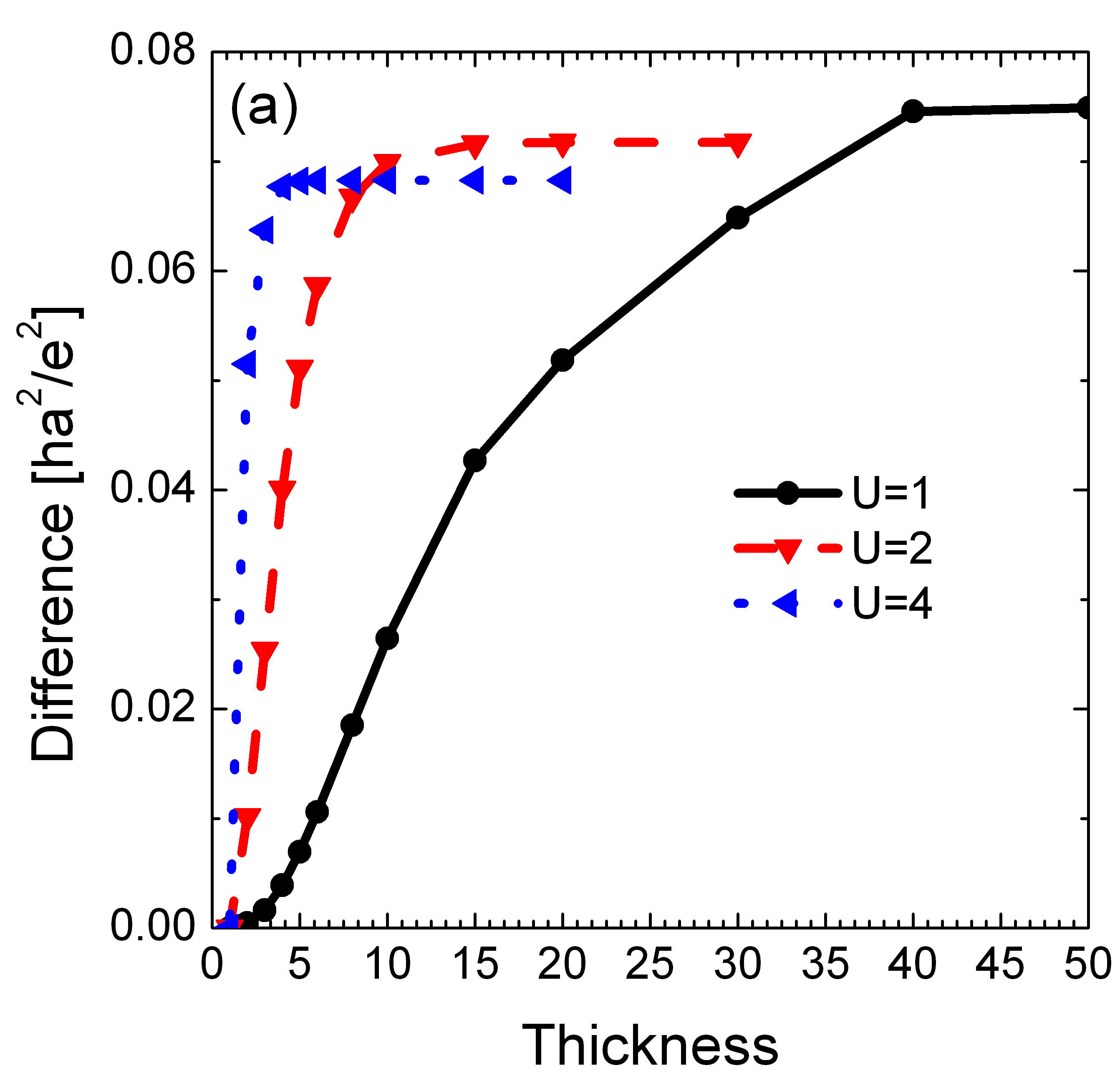}
			\includegraphics[width=80mm]{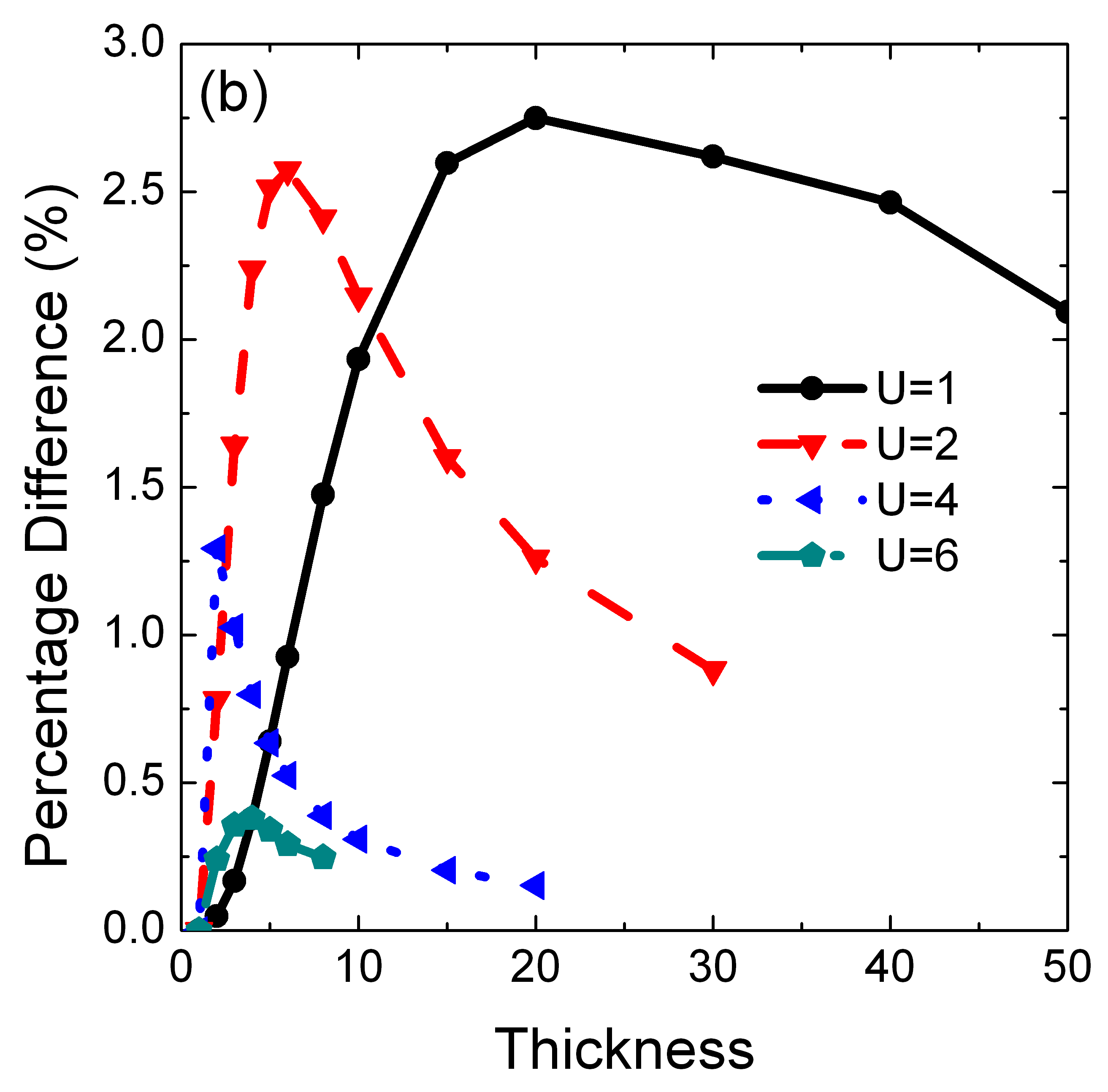}
	\caption{(Color online) Difference between vertex-corrected and non vertex-corrected resistance values as a function of barrier thickness. Panel (a) shows the difference between values of the resistance for various interaction strengths, $U$, showing that the difference grows with an increase in barrier width until it plateaus at a saturation value that decreases as the interaction strengths increases towards the metal-insulator transition value. Panel (b) shows the percentage difference between vertex-corrected and non vertex-corrected resistance values $[(R_{no\  vertex}-R_{vertex})/R_{no\  vertex}\times100]$. Note how the lower the interaction strength is, the thicker the barrier must be before the saturation is reached, indicated by the flat region in panel (a).}
	\label{fig:diff}
\end{figure} 
	In Fig. \ref{fig:diff}, we plot the difference between the vertex-corrected and non vertex-corrected terms to examine the effects of vertex corrections on various thicknesses. As expected the vertex corrections are relatively small and as the thickness increases the vertex corrections saturate in magnitude. This saturation is because as the thickness increases the barrier approaches the bulk limit, and in the bulk the vertex corrections vanish. Therefore the effect of vertex corrections lies primarily in the interface region and they disappear as you move away from interface and into the ``bulk'', part of the barrier, which dominates for thicker barriers.
	
\begin{center}
\textbf{IV. CONCLUSIONS}
\end{center}

	In this work, we used inhomogeneous DMFT to calculate the electronic \textit{dc} charge transport, including the effect of vertex corrections in the Falicov-Kimball model. We used a Kubo-Greenwood formula to derive the \textit{dc} conductivity matrix, using the exact expression for the vertex corrections, known for the Falicov-Kimball model. Although we only applied the vertex corrections to the Falicov-Kimball model, the results up to Eq.~(\ref{eq:presub}) hold for any general model. Therefore, the theory can be extended to other models if the corresponding irreducible vertex can be found. We can easily modify the model by adding mean-field-like interactions such as Zeeman splitting for magnetic systems, and use whatever impurity solver is desired to calculate the local Green's functions on each plane. 

	In addition to changing the general model, there are a number of transport effects and model modifications that could yield interesting results beyond what is presented in this paper. We showed that vertex corrections play a small role in \textit{dc} conductivity but did not investigate transport outside the \textit{dc} limit. We also neglected charge reconstruction\cite{Nikolic, Ling} at the interface and by performing the calculations off of half-filling one could introduce temperature dependence to the chemical potential. Finally, by breaking the particle-hole symmetry one could investigate the effect of vertex-corrections in thermal transport. But in all cases, we expect the vertex corrections to be small for the same reasons they are small here. 
	
	By varying the barrier thickness, we see the effects the vertex corrections play as we go from the single barrier layer limit to the bulk limit. In Figure \ref{fig:diff}, we see that the vertex corrections make a small relative change to the \textit{dc} conductivity saturates as the barrier thickness increases towards the bulk. We also see that vertex corrections become relatively smaller as the Falicov-Kimball interaction strength increases. Hence, the effect of the vertex corrections are small enough that they probably do not need to be included in an inhomogeneous DMFT calculation for longitudinal transport. Since the thermal transport arises from the Jonson-Mahan Theorem\cite{FreeZS} vertex corrections also should not affect the thermal transport much.

\acknowledgments
We acknowledge the support of the National Science Foundation through grants DMR-0705266 and DMR-1006605. We would also like to acknowledge useful conversations with Joerg Schmalian and V. Zlati\'{c}. This work was initiated from a discussion at the Aspen Center for Physics in 2008.   

\end{document}